# Crystallization-Arrested Viscoelastic Phase Separation in Semiconducting Polymer Gels

*Jing He†, Xiaoqing Kong†, Yuhao Wang §, Michael Delaney†, Dilhan M. Kalyon†,‡ and Stephanie S. Lee†,\**

†Department of Chemical Engineering and Materials Science, Stevens Institute of Technology, Hoboken, NJ 07030, USA

§Department of Biomedical Engineering, Stevens Institute of Technology, Hoboken, NJ 07030, USA

‡Highly Filled Materials Institute, Stevens Institute of Technology, Hoboken, NJ 07030, USA

ABSTRACT: Through a combination of rheological characterization and temperature-variable imaging methods, a novel gelation pathway in dilute solutions of a semiconducting polymer to achieve interconnected, crystalline networks with hierarchical porosity is reported. Upon rapid cooling, solutions of regioregular poly(3-hexylthiophene) (RR-P3HT) in ortho-dichlorobenzene formed thermoreversible gels. Temperature-variable confocal microscopy revealed cooling-induced structural rearrangement to progress through viscoelastic phase separation. The phase separation process arrested prematurely during the formation of micron-sized solvent-rich

"holes" within the RR-P3HT matrix due to intrachain crystallization. Cryogen-based scanning electron microscopy of RR-P3HT gels revealed the existence of an interfibrillar network exhibiting nano-sized pores. Remarkably, these networks formed to equal gel strengths when a third component, either small molecule phenyl-$C_{61}$-butyric acid methyl ester (PCBM) or non-crystallizing regiorandom (Rra)-P3HT, was added to the solution. Organic solar cells in which the active layers were deposited from phase-separated solutions displayed 45% higher efficiency compared to reference cells.

1. **Introduction**

Viscoelastic phase separation (VPS), a thermodynamic phenomenon unique to systems in which there exists a dynamic asymmetry between the components, is a promising avenue to direct the self-assembly of materials into hierarchical structures.[1] Unlike classical phase separation, VPS initiates through the formation of droplets of the fast-moving component, e.g. a small-molecule solvent, within a matrix of the slow-moving component, e.g. a long-chain polymer.[2-4] If attractive molecular interactions are sufficiently strong in the slow-moving phase, the system will form a transient gel. Over time, the droplets of the fast-moving phase grow and coalesce to form the majority phase. However, if VPS is arrested in the transient gel phase, the formation of stable cellular and network structures is possible.

The ability to arrest VPS during the early stages will afford control over the morphology of porous materials for applications ranging from membrane filtration to plastic foam manufacturing, and is likely responsible for naturally-occurring network structures found in, for example, magma[3] and foods[5]. By evacuating the solvent during the early stages of VPS, for example, porous scaffolds for tissue engineering were formed from polymer blends exhibiting dynamic asymmetry between the two components.[6] Crystallization-arrested VPS has also been observed in protein solutions at high concentrations, resulting in the formation of stable gels.[7] Recently, Tanaka and coworkers reported crystallization-arrested VPS in colloidal suspensions to form "crystal gels."[8] This structure was achieved inducing VPS below the melting point of the colloid crystals such that nucleation and crystal growth occur during the formation of perocolated networks of the liquid colloidal phase.

The formation of porous crystalline networks via VPS presents a promising, scalable route to achieve optimized morphologies in the active layers of organic solar cells (OSCs). In these devices, the photoactive layer is typically deposited from a solution comprising a blend of an electron-donating polymer and electron-accepting small-molecule. The composition of these solutions with dynamically-asymmetric components are typical of those that have been demonstrated to undergo VPS in the past. For the application of light energy harvesting, porous polymer network exhibiting large interfacial surface area with the electron acceptor promote efficient exciton dissociation. Furthermore, polymer crystallization within these percolated networks would facilitate efficient hole transport to the anode. Given that morphological control over OSC active layers remains a critical challenge[9], VPS is an intriguing avenue towards robust control over solution structures prior to active layer deposition.

To this end, we herein demonstrate VPS in electron-donating poly(3-hexlythiophene) (RR-P3HT) solutions upon rapid cooling to form stable, crystalline RR-P3HT gel networks with hierarchical porosity that promotes phase separation with electron acceptor phenyl-$C_{61}$-butyric acid methyl ester (PCBM). A combination of rheological characterization and temperature-dependent fluorescence microscopy revealed that VPS is arrested during solvent hole formation due to interchain crystallization of RR-P3HT. This crystallization-induced gelation during VPS was found to be thermoreversible and remarkably insensitive to the presence of both PCBM and non-crystallizing regiorandom-P3HT (Rra-P3HT). Once formed in solution, these semi-crystalline RR-P3HT networks can be transferred to OSC device platforms via doctor blading, a method compatible with continuous processing methods. OSCs comprising cooled photoactive layers displayed 45% higher efficiencies compared to those comprising uncooled photoactive layers. Through the discovery of a novel gelation mechanism in this extensively studied system, we thus present a viable strategy to control the photoactive layer morphology *prior* to film deposition for improved light conversion efficiency in scalable OSCs.

2. **Experimental Section**

*Materials*: Two different types of P3HT were used as received from Rieke Metals (Lincoln, NE): regioregular P3HT (RR-P3HT) with $M_W$=50-70 kDa, regioregularity=91% and $M_W$=70-90 kDa regiorandom P3HT (Rra-P3HT). PCBM was used as received from Solarmer energy Inc (Irwindale, CA). o-DCB was used as received from Sigma-Aldrich.

*Sample preparation*: Solutions of P3HT were prepared by dissolving P3HT in o-DCB at 500 rpm and 60 °C on a magnetic stirrer hotplate overnight. Total P3HT concentrations were varied between 12.5 mg/mL, 25 mg/mL, 37.5 mg/mL, and 50 mg/mL. Solutions of blends of RR-P3HT and Rra-P3HT were prepared by weight ratios of 1:1 or 2:1.

*Rheology*: An Advanced Rheometric Expansion System (ARES) rheometer available from TA Instruments of New Castle, DE, was used in conjunction with a force rebalance transducer 0.2K-FRTN1 and stainless steel parallel disks with 8-25 mm diameter for small-amplitude oscillatory shearing experiments. The torque accuracy of the transducer is ±0.02 g-cm. The actuator of the ARES is a DC servo motor with a shaft supported by an air bearing with an angular displacement range of $5 \times 10^{-6}$ to 0.5 rad, and angular frequency range of $1 \times 10^{-5}$ to 100 rad/s. It has an angular velocity range of $1 \times 10^{-6}$ to 200 rad/s. The top disk was attached to the upper fixture which was connected to the torque and normal force transducers and the bottom disk was attached to the bottom fixture of the rheometer, which was coupled to the motor. The rheometer was equipped with an environmental control chamber which can operate from -150 °C to 600 °C with a ramp rate from 0.1 to 50 °C/min. During oscillatory shearing the shear strain $\gamma$, varies sinusoidally with time, $t$, at a frequency of $\omega$, i.e., $\gamma(t) = \gamma^0 \sin(\omega t)$ where $\gamma^0$ is the strain amplitude. The shear stress, $\tau(t)$ response of the fluid to the imposed oscillatory deformation consists of two contributions associated with the energy stored as elastic energy and energy dissipated as heat, i.e., $\tau(t) = G'(\omega)\gamma^0 \sin(\omega t) + G''(\omega)\gamma^0 \cos(\omega t)$ The storage modulus, $G'(\omega)$ the loss modulus, $G''(\omega)$, also define the magnitude of complex viscosity, $|\eta^*| = \sqrt{(G'/\omega)^2 + (G''/\omega)^2}$, and $\tan\delta = G''/G'$. In the linear viscoelastic region all dynamic properties are independent of the strain amplitude, $\gamma^0$.

The solution was kept stirring on the hotplate at 60 °C prior to loading into the gap between two parallel disks. The temperature was dropped to targeted sub-ambient temperatures, -5 °C, -10 °C, and -15 °C, within 1 minute after loading the solution in between parallel disks at room temperature. A thin layer of Fomblin oil was applied onto the free surface of the sandwiched sample to avoid solvent loss.[10] For each test, a fresh sample was loaded onto the rheometer to avoid any thermal and mechanical history effects, unless indicated specifically. Special care was taken to probe the wall slip effect by systematically varying the surface to volume ratios by systematic changes in the testing gap between 0.5 mm, 0.75mm, and 1mm.[11] In order to characterize the sol-gel transition kinetics, the oscillatory shear data was collected first as a function of time at constant $\gamma^0$ and $\omega$. The time sweep tests were stopped upon recording of ultimate plateau values of $G'$ and $G''$ that were sustained for at least 10 minutes. This indicates that the gel formed is at steady-state and the solvent loss is negligible. Following tests were carried out by collecting $G'$ and $G''$ values as a function of frequency at a constant $\gamma^0$ within the linear viscoelastic region. Temperature ramp tests were carried out to collect oscillatory shear data as a function of temperature at constant $\gamma^0$ and $\omega$.

*Cryo-scanning electron microscopy*: 2 µL solutions were kept in aluminum sample holders and cooled to -5 °C or -15 °C for 30 min to induce gelation based on previous time sweep rheological data. The gelled samples were then soaked in a liquid nitrogen bath at -199 °C to keep the microstructures intact during transfer for cryo-SEM imaging. The samples were then placed into a humidity-free cryo-transfer system (Leica EM VCT100) before SEM imaging. In order to reveal the P3HT gel structure, solvent sublimation was carried out at -80°C under a vacuum of $10^{-7}$ mbar for 10 minutes to remove the solvent and condensed ice on the sample surfaces. The samples were then cooled down to -130 °C again for SEM imaging. SEM was carried out using a Zeiss Auriga Dual-Beam FIB-SEM (Carl Zeiss Microscopy) at an accelerating voltage of 5 kV using an Everhart-Thornley secondary electron detector.

*Confocal laser scanning microscopy:* A Nikon Ti-E inverted microscope along with C1 CLSM system was used to characterize morphological development of P3HT solutions during cooling. The solution was encapsulated inside a rectangular glass chamber with a wall thickness of 200 µm, with another glass chamber filled with dry air attached below as a heat insulation layer. A Thermo Scientific NESLAB RTE 7 fluid bath was employed to control the temperature. Nitrogen gas was continuously purging the outer surface of the glass chamber to avoid water condensation under sub-ambient temperatures. Heat flux sensor and thermocouples were employed to determine the true temperature distribution within the sample (refer to Figure S1 for a schematic of the setup and temperature distribution). P3HT was excited at a wavelength of 488 nm, and the emitted signal at a wavelength of 590 ± 30 nm was collected.

*Solar cell device fabrication and characterization*: Patterned ITO coated glass substrates (15 Ω sq$^{-1}$, Xinyang Technology Co.) were cleaned by sonication with acetone, deionized water, isopropanol, for 10 min each sequentially, followed by a 15 min $O_2$ plasma treatment. Thin layers (80 nm) of PEDOT:PSS (Clevios F HC, Heraeus) were spin coated onto cleaned ITO substrates at 5000 rpm for 90 s, and then annealed at 150 °C for 30 min in air. RR-P3HT/PCBM solutions that had been cooled to -5 °C for various amounts of time were then deposited on top of the PEDOT:PSS layers by blade coating. The devices were then transferred to a nitrogen glove box for thermal annealing at 150 °C for 30 min. Finally, 100 nm Al electrodes were deposited on top of the photoactive layers with a thermal evaporator (Covap, Angstrom Engineering Inc.) through shadow mask. The device area was defined to be 0.046 cm$^2$. Current density versus voltage curves were collected in a $N_2$ atmosphere using a Keithley 2636B sourcemeter under simulated AM 1.5G irradiation (100 mW cm$^{-2}$). The intensity of the xenon arc lamp based simulator (Model 11002, Abet Technologies Inc.) was calibrated by using a monocrystalline Si reference cell.

## 3. Results and discussion

Regioregular-poly(3-hexylthiophene) (RR-P3HT), the most intensely studied semiconducting polymer for OSCs,[12] is known to undergo gelation in organic solvents. In the first study of the gelation of RR-P3HT solutions dissolved in xylene, a poor solvent for RR-P3HT, a two-step gelation mechanism involving a coil-to-rod transition followed by crystallization of the rods was proposed.[13] Subsequent rheological and absorption measurements supported this mechanism of RR-P3HT aggregation via π-π interactions, followed by physical linking of the aggregates to form a gel network.[14] A similar pathway was observed for RR-P3HT dissolved in p-xylene, toluene and benzene, using a combination of rheology and electrical conductivity characterization to determine the extent of percolation.[10] Small-angle x-ray scattering experiments of RR-P3HT/xylene gels later revealed the existence of RR-P3HT nanowhiskers.[15] These nanowhiskers were found to be only semicrystalline, likely due to restrictions imposed by extensive interchain interactions. Rheological tests to measure gel strength revealed that the gelation process can be inhibited by the presence of phenyl-$C_{61}$-butyric acid methyl ester (PCBM), a common electron acceptor co-deposited with RR-P3HT to form photoactive layers.[16]

Previous literature reports have found that cooling RR-P3HT solutions can induce aggregation by lowering the solubility of RR-P3HT in the solvent.[17-20] To monitor the cooling-induced gelation of RR-P3HT solutions in o-DCB, we performed a series of oscillatory shear tests as a function of P3HT concentration, frequency, and temperature. Figure 1a displays the storage modulus, $G'$, of RR-P3HT solutions cooled to -5 °C from room temperature within 30 seconds and kept at -5 °C for the characterization of the dynamic properties as a function of time. The experiments were carried out at RR-P3HT concentrations of 12.5, 25, and 50 mg/ml. Linear viscoelastic moduli were collected as a function of time at -5 °C under small-amplitude oscillatory shear (SAOS) conditions, with a strain amplitude, $\gamma^0$, and frequency, $\varpi$, of 1% and 1 rad/s, respectively (Figure S2). As displayed in Figure 1, all three solutions initially exhibited a large increase in $G'$, followed by a monotonic increase to a plateau, i.e., reached a concentration-dependent steady-state value of $G'$, denoted as $G'_{max}$. The value of $G'_{max}$ was found to increase from 1 to 50 kPa as the RR-P3HT concentration in o-DCB was increased from 12.5 to 50 mg/ml. The loss moduli, $G''$, values collected as a function of time for all three concentrations of the solutions were negligible in comparison to the values of $G'$, i.e., the $G''$ values were only in the 10 - 1000 Pa, thus, were at least 50 times smaller than the corresponding storage moduli.

In addition to affecting the final gel strength, the concentration of RR-P3HT in solution was also found to affect the rate of gelation. A characteristic gelation time, $t_c$, was defined as the time at which the maximum value of $\mathrm{d}G'(t)/\mathrm{d}t$ normalized to $G'_{max}$ was observed to track the gelation rate (Figure S3). With increasing RR-P3HT concentration from 12.5 to 50 mg/mL, $t_c$ decreased from 2000 to 200 s, indicating a faster rate of gelation with increasing RR-P3HT concentration.

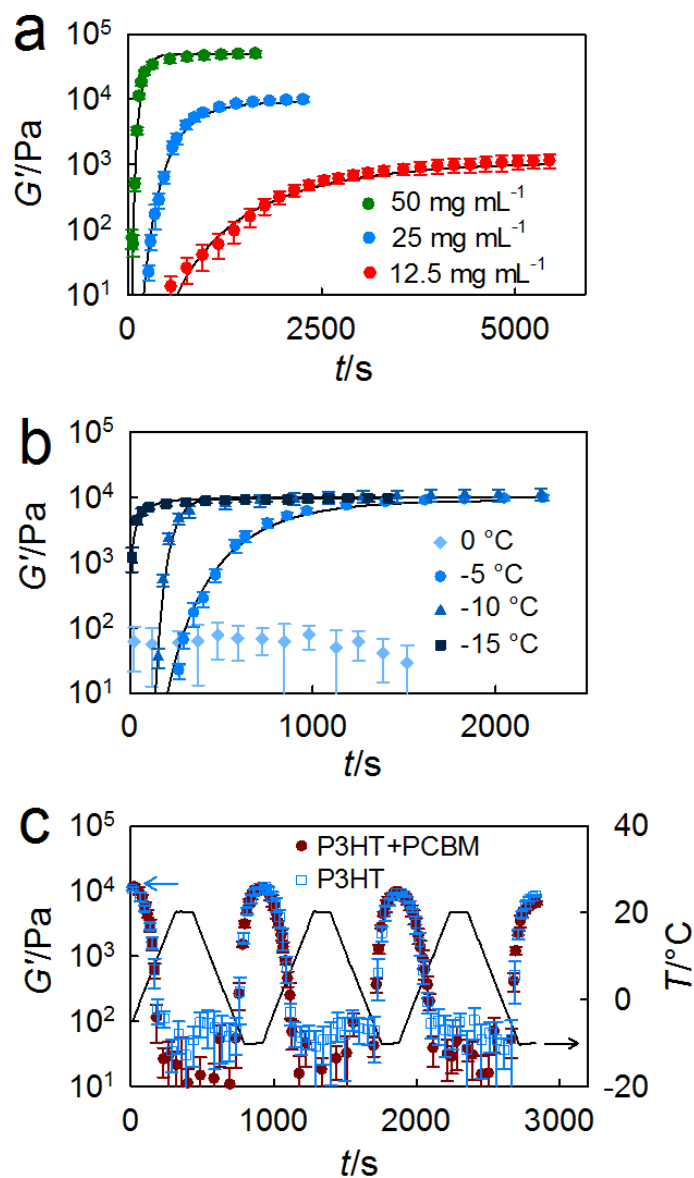

**Figure 1.** a) Storage modulus, $G'$, versus time, $t$, of RR-P3HT solutions in o-DCB with concentrations of 12.5, 25, and 50 mg/mL cooled from room temperature to -5 °C. Lines provided to guide the eye. b) Storage modulus, $G'$, versus time, $t$, of 25 mg/mL RR-P3HT solutions cooled to 0, -5, -10, and -15 °C. c) $G'$ of a pre-cooled 25 mg/mL RR-P3HT with and without 25 mg/mL PCBM versus $t$ during temperature cycling between -10 and 20 °C.

To probe the effect of temperature on the development of the rate of gelation and gel strength, time sweeps were carried out on fresh 25 mg/ml RR-P3HT solutions in o-DCB which were cooled to temperatures in the range of 0 to -15 °C. All of the targeted temperatures were reached within 30 seconds after the environmental chamber was closed and temperature control was switched on. As displayed in Figure 1b, $G'$ of a 25 mg/ml RR-P3HT solution in o-DCB cooled to 0 °C remained negligible for at least 1500 s, indicating that the solution remained as a viscous fluid during these 1500 s. For 25 mg/mL RR-P3HT solutions cooled to -5 to -15 °C, on the other hand, a dramatic increase in G' from negligible values to a $G'_{max} = 10^4 \, \text{Pa}$ was observed. The $G'_{max}$ values of RR-P3HT, i.e., the gel strengths, were independent of temperature in the -5 to -15 °C range (Figure S4). However, the characteristic gelation time, $t_c$, decreased from 800 to 250 s as the temperature decreased from -5 to -15 °C, reflecting a significant increase of the rate of gelation with decreasing temperature (Figure S5). Evidence of a sharp transition from RR-P3HT unimers to RR-P3HT aggregates in o-DCB between 0 and -5 °C was recently observed using Raman spectroscopy.[21] This transition was attributed to the solution concentration coinciding with the solubility limit of RR-P3HT, which decreases with decreasing temperature.[22]

After stable gel formation, the cooled RR-P3HT solutions were subjected to frequency sweeps in the linear viscoelastic region (Figure S6). For all concentrations of RR-P3HT, $G'$ values were constant at their respective $G'_{max}$ values over a frequency range of 0.1 to 100 rad/s. The observations that $G'$ and $G''$ are independent of frequency and that the value of $\tan\delta = G''/G'$ is smaller than 0.1 in this range, i.e., $G'(\varpi) \gg G''(\varpi)$, are both hallmarks of gel-like behavior.[23] The observed $G'(\varpi) \gg G''(\varpi)$ ($\tan\delta = G''/G'$ is smaller than 0.1) indicates that relatively long relaxation times prevail, i.e., a precursor of solid-like behavior (Figure S7). The time-dependent increase of the storage modulus describes the transformation of the RR-P3HT solution from a fluid to a gel, i.e., "gelation process for RR-P3HT."

We also performed oscillatory shear experiments during temperature cycling between 20 $^{\circ}$C and -10 $^{\circ}$C on 25 mg/ml RR-P3HT solutions in o-DCB that were pre-cooled to -5 °C for 30 minutes to examine the reversibility of the observed sol-gel transition. Figure 1c displays the temperature and storage modulus of the solution versus time. $G'$ tracked closely to the temperature profile during cycling, displaying values of $<100\,\text{Pa}$ and $10^4\,\text{Pa}$ at 20 $^{\circ}$C and -10 $^{\circ}$C, respectively. Interestingly, the characteristic time for gelation during temperature cycling was 5 times faster compared to the initial gelation rate of a fresh RR-P3HT solution cooled from room temperature. This observation suggests that some "memory" of the gel structure persists during heating to 20 $^{\circ}$C, even though the solution behavior approaches the behavior of a purely viscous fluid at 20 $^{\circ}$C. Oh and coworkers studied the growth of P3HT nanofibrils in solutions by a cycle of cooling and heating, concluding that such "memory" effect is due to the crystallinity hysteresis upon temperature cycling.[19] Newbloom and coworkers likewise found persistence of the P3HT fiber network after dissolution of the gel phase.[24]

For bulk-heterojunction OSC applications, RR-P3HT is most commonly co-dissolved in solution with phenyl-$C_{61}$-butyric acid methyl ester (PCBM), which is then spun cast onto device platforms to form the photoactive layer. The presence of PCBM has been found to significantly disrupt the crystallization of RR-P3HT when the two compounds are co-deposited in this manner.[25, 26] To examine the effect of PCBM on the thermoreversible gelation of RR-P3HT, the linear viscoelastic material functions of a solution comprising 25 mg/ml of PCBM co-dissolved with 25 mg/ml RR-P3HT in o-DCB were characterized. Surprisingly, the presence of PCBM did not significantly alter the thermoreversible gelation behavior of RR-P3HT, with no observable change in the characteristic gelation time or gel strength (Figure 1c). This finding contradicts a previous report which suggested that both the rate of gelation and steady-state gel strength of 80 mg/ml RR-P3HT dissolved in o-DCB at 5 °C significantly decreased in the presence of PCBM.[16] We hypothesize that this discrepancy is due to differences in the gelation mechanism in different temperature regimes. RR-P3HT solutions form gels at room temperature over a period of days via the crystallization-driven self-assembly of RR-P3HT chains into nanowhiskers, which then interconnect to form a network.[14, 15, 27-29] In contrast, we observe gelation to occur on the order of minutes in RR-P3HT solutions cooled rapidly to sub-ambient temperatures, which likely proceeds via an alternate gelation pathway.

To examine the role of crystallization on rapid RR-P3HT gelation upon cooling, we examined the behavior of solutions comprising non-crystallizing regiorandom (Rra)-P3HT of similar molecular weight to that of the RR-P3HT used in these studies. Time sweeps of 25 mg/ml RRa-P3HT solutions in o-DCB revealed that RRa-P3HT does not form a gel upon cooling to -5 °C. As displayed in Figure 2a, the storage modulus of the solution comprising pure Rra-P3HT at a concentration of 25 mg/ml remained below 10 Pa upon cooling. The observation that non-crystallizing Rra-P3HT does not form a gel thus indicates that intrachain crystallization, not chain entanglements, form the junctions in RR-P3HT gels.

In contrast to pure Rra-P3HT solutions, solutions comprising blends of RR-P3HT and Rra-P3HT with concentration ratios from 1:0.5 to 1:2 and total P3HT concentrations up to 50 mg/ml in o-DCB still exhibited gel formation upon cooling to -5 °C. Surprisingly, $G'_{max}$ was found to depend solely on the concentration of RR-P3HT and was independent of the Rra-P3HT concentration in the blended solutions (Figure 2b). These results indicate that the presence of Rra-P3HT does not disrupt the crystallization of RR-P3HT in solution, nor does its presence contribute to the gel strength despite an increase in the total polymer content.

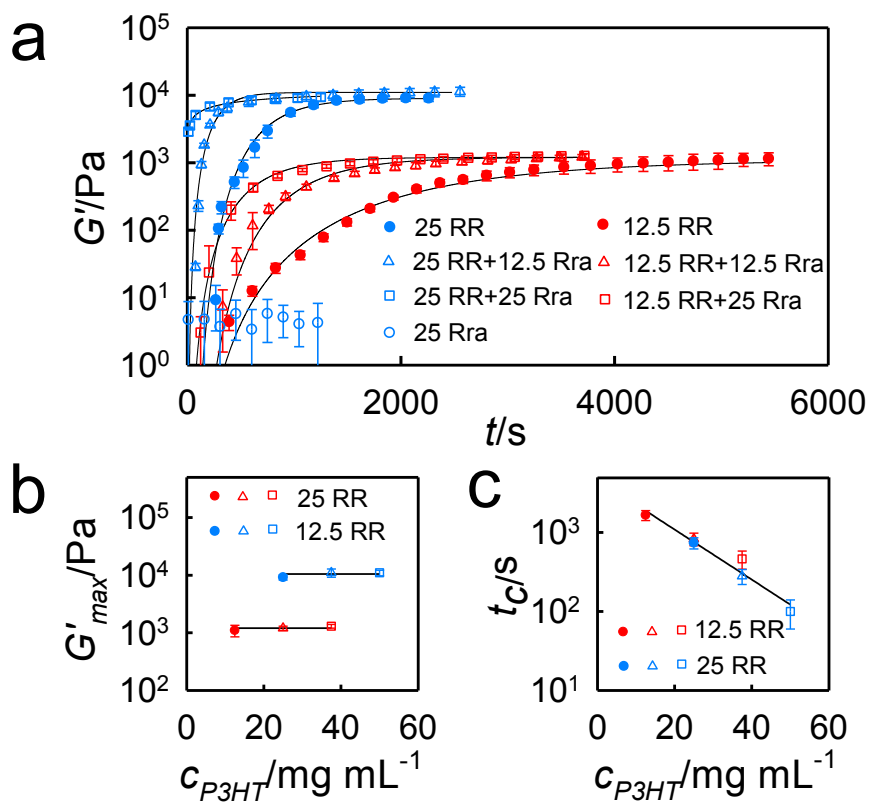

**Figure 2.** a) $G'$ versus time of solutions comprising blends of RR- and Rra-P3HT in o-DCB with concentrations of 12.5 and 25 mg/mL, cooled from room temperature to -5 ºC. b) Steady-state storage modulus, $G'_{max}$, of the solutions versus the total concentration of P3HT. c) Characteristic gelation time of the solutions, $t_c$, versus the total concentration of P3HT.

However, although the gel strength is not affected, the rate of gelation was found to strongly depend on the total P3HT content, as displayed in Figure 2c. The characteristic gelation time, $t_c$, decreased significantly from 2000 s to 100 s as the P3HT content (RR- plus Rra-P3HT) increased from 12.5 mg/mL to 50 mg/mL, independent of the specific ratio of Rra-P3HT to RR-P3HT. The presence of Rra-P3HT thus accelerates the rate of gelation, which is attributed to network formation of RR-P3HT. Previous findings in the literature are in disagreement on the ability of Rra-P3HT to disrupt the crystallization of RR-P3HT in blends. For example, a two order-of-magnitude decrease in hole mobility upon addition of Rra-P3HT to RR-P3HT films acting as active layers in field-effect transistors was attributed to decreased RR-P3HT crystallization.[30] On the other hand, unfavorable interactions between Rra-P3HT and RR-P3HT were found to lead to vertical phase separation between the two polymers in transistor active layers.[31] Devices comprising these vertically-segregated blends displayed no significant decrease in performance compared to those comprising pure RR-P3HT in the active layer. To the best of our knowledge, the acceleration of RR-P3HT crystallization in the presence of Rra-P3HT has not yet been reported.

Combined, our observations that the presence of Rra-P3HT 1) does not affect the final gel strength and 2) increases the rate of gelation provide direct evidence for a two-step gelation mechanism in these semi-rigid polymer solutions. In agreement with our finding that the presence of PCBM does not affect the kinetics or extent of RR-P3HT gelation, we hypothesize that cooling the solutions below the solubility limit of RR-P3HT first induces phase separation of RR-P3HT into polymer-rich and polymer-poor domains within the solution. This phase separation is followed by interchain crystallization of the RR-P3HT phase into nanofibrils that form the gel network. In the absence of phase separation, we would expect the presence of Rra-P3HT chains to impede RR-P3HT crystallization. The characteristic time for gelation, $t_c$, is thus a measure of the rate of phase separation to form RR-P3HT-rich domains and $G'_{max}$ is a measure of the number density of crystalline affine junction points, crosslinks, formed in the gel network.

Direct high-resolution imaging of P3HT gels to confirm the network morphology is exceedingly difficult because of the presence of organic solvents that are generally incompatible with vacuum-based imaging techniques. Furthermore, processing methods that rely on the removal of the solvent for subsequent imaging, such as heating or slow solvent evaporation, unavoidably alter the gel morphology. Using a novel cryogen-based imaging method pioneered by the Libera group,[32] we collected high-resolution SEM images of cooled P3HT gels in partially solvated environments to capture morphological features on the nanometer length scale. As detailed in the Experimental Section, 25 mg/ml P3HT solutions were cooled to -5 $^{o}$C for 30 min to induce gelation. The gels were then flash-frozen in liquid nitrogen, followed by partial solvent sublimation at a pressure of $1 \times 10^{-7}$ mbar and temperature of -80 $^{o}$C for 10 min. Longer sublimation times to completely remove the solvent led to collapse of the gel structure, as confirmed by SEM imaging (Figure S8).

Figure 3a displays an SEM image of a partially-solvated RR-P3HT gel. As observed in the figure, a nanofibrous interconnected network with pore sizes on the tens to hundreds of nanometers length scale is observed. This morphology is consistent with our proposed mechanism of interchain polymer crystallization during cooling and is reminiscent of structures observed in freeze-dried P3HT gel samples.[33] We observe fiber diameters that are comparable to those previously reported.[34, 35] For reference, similarly processed samples comprising Rra-P3HT appeared glassy and did not display a fibrous network due to the inability of Rra-P3HT to crystallize (Figure 3b).

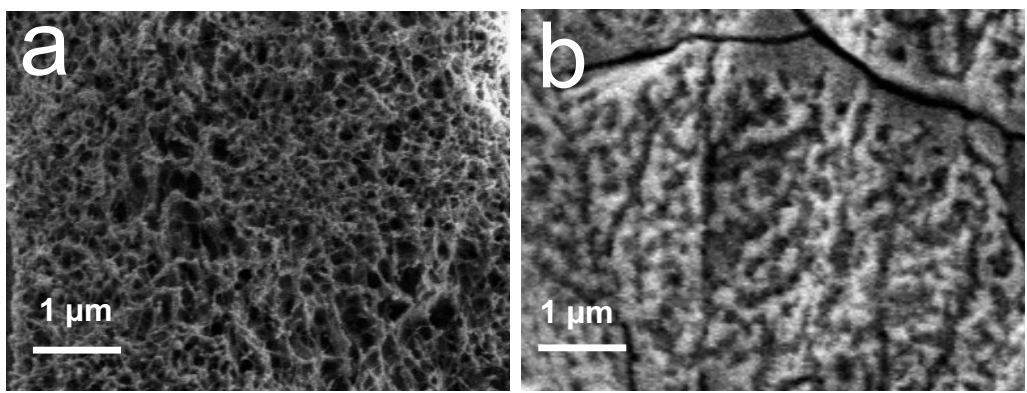

**Figure 3.** Cryogen-based SEM images of partially-desolvated samples comprising a) RR-P3HT and b) Rra-P3HT.

To monitor morphological development of P3HT solutions in real-time during cooling, we exploited the photoluminescence (PL) properties of P3HT. When excited at a wavelength of 488 nm, P3HT emits a PL signal between 500 nm and 700 nm, with a maximum PL signal around 570 nm.[36] We employed temperature-variable CLSM (confocal laser scanning microscopy) to observe the gelation process and gain insights into the gelation mechanism in our system. The PL signal of RR- and Rra- P3HT solutions in o-DCB at room temperature and cooled to -5 °C under 488-nm laser excitation was mapped. As displayed in Figure 4a, the fluorescence signal collected on a 25 mg/ml RR-P3HT solution in o-DCB at room temperature was spatially uniform, indicating that the polymer was well-dissolved in the solvent. Upon cooling to -5 °C, dark "holes" with diameters ranging from 1 to 5 μm in the fluorescence micrograph appeared (Figure 4b). These solvent-rich holes, a hallmark of the early stages of viscoelastic phase separation, form due to dynamic asymmetry between the solvent molecules and polymer chains.[37] The natural progression of viscoelastic phase separation is for the solvent holes to grow and coalesce, eventually forming the majority phase. Interestingly, the holes in cooled RR-P3HT solutions persisted until the solutions were re-heated to room temperature. Our rheological and fluorescence microscopy experiments indicate that these porous gels are stable, with no change in gel properties or morphology after 20 min. We thus hypothesize that interchain RR-P3HT crystallization arrests viscoelastic phase separation in the early stages of solvent hole formation, as illustrated in Figure 4c.

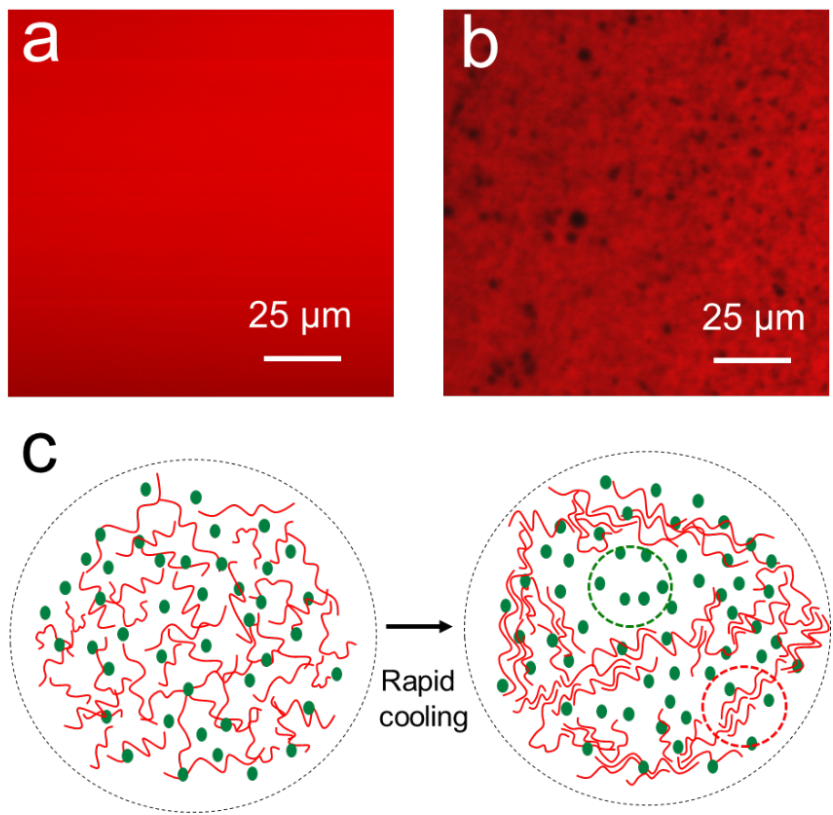

**Figure 4.** Confocal laser scanning microscopy images of a 25 mg/mL RR-P3HT solution in o-DCB at a) 20 °C and b) at -5 °C. c) Illustration of the RR-P3HT solution before and after cooling-induced viscoelastic phase separation. A solvent hole and interchain crystal are highlighted by green and red dashed circles, respectively.

The formation of porous "crystal gels" via crystallization-arrested viscoelastic phase separation has previously been observed in supercooled colloidal suspensions.[8, 38] Monitoring the process at the single-particle level, Tanaka and coworkers recently proposed a set of requirements for observing this phenomenon, including a driving force for phase separation below the melting point of the colloidal crystal phase and a low energy barrier to crystallization.[8] In our system, we induce phase separation by rapidly cooling the solution to lower the solubility of P3HT in the solvent. Once the solubility of P3HT in o-DCB falls below the solution concentration, phase separation is induced. Because o-DCB is a relatively good solvent for P3HT,[39] we expect the chains to adopt large end-to-end distances, which in turn facilitates interchain crystallization. Given the ability to tune the driving force for interchain crystallization by varying the degree of regioregularity[40, 41] and the solvent quality,[42-44] this model system will open new avenues to explore factors governing crystallization-arrested viscoelastic phase separation in dilute polymer solutions.

Network structures have critical applications in fields ranging from membrane filtration to food processing.[1] In the field of solution-processable OSCs, the reproducible formation of interconnected, crystalline networks of semiconducting polymers can significantly affect device performance by providing a continuous, high-mobility network for efficient charge transport and large interfacial area with the electron-accepting (or donating) phase for exciton dissociation. Critically, controlling the solution structure prior to deposition will enable the continuous processing of high-performance OSC photoactive layers in a scalable manner. As a demonstration of morphological control in OSC photoactive layers via solution-phase gelation to improve OSC device performance, we fabricated OSCs comprising doctor-bladed RR-P3HT/PCBM solutions cooled to -5 °C for varying amounts of time (see Experimental section for fabrication details). Figure 5a displays characteristic J-V curves for OSCs comprising RR-P3HT/PCBM photoactive layers deposited without solution cooling and with 300 s of solution cooling to -5 °C. The short circuit current, $J_{sc}$, open circuit voltage, $V_{oc}$, and fill factor, $FF$, of the OSC comprising a photoactive layer deposited from a pre-cooled solution were all found to be higher compared to the reference device, resulting in an overall improvement in the light conversion efficiency from 2.7 ± 0.2% to 3.9 ± 0.2%.

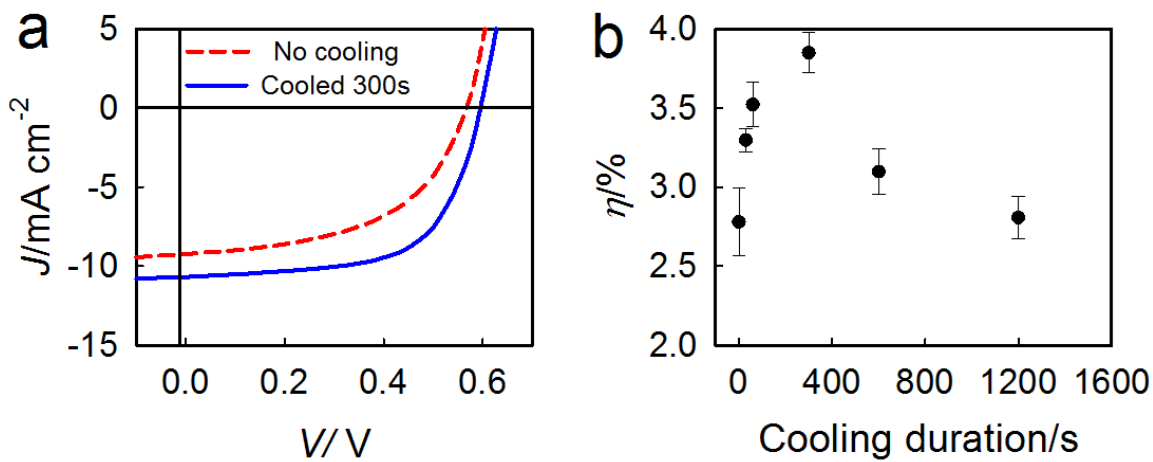

**Figure 5**. a) Representative J-V curves of OSCs comprising photoactive layers blade-coated from an uncooled solution and a solution cooled to -5 °C for 300 s. b) Efficiency of OSCs versus the cooling duration of the RR-P3HT/PCBM solution prior to blade coating. Error bars represent the standard deviation in efficiencies for three devices.

Figure 5b displays the light conversion efficiency, $\eta$, as a function of the solution cooling duration at -5 °C prior to blade coating. Interestingly, the maximum efficiency is observed for a cooling duration of 300 s. Upon longer cooling times, the efficiency decreased. Based on our rheological measurements displayed in Figure 1, gel formation completes by ~1000 s for solutions comprising 25 mg/ml RR-P3HT. We speculate that this decrease in efficiency is related to the extent of phase separation between RR-P3HT and PCBM at different points in the cooling process. As a small molecule, PCBM is not expected to display significant dynamic asymmetry with the solvent. PCBM is thus present within the solvent holes that form during viscoelastic phase separation between the solvent and RR-P3HT. As these holes grow, likewise do the PCBM-rich domains that form upon solvent removal during blade coating. It is well-established that while some degree of phase separation between RR-P3HT and PCBM improves device efficiency by increasing the number of continuous pathways in the RR-P3HT and PCBM phases to the anode and cathode, respectively, extensive phase separation results in a decrease in interfacial area for exciton dissociation and thus lower device efficiency.[45-47] In our proposed method, we can control the extent of phase separation between RR-P3HT and PCBM by stopping the gelation process at different time points.

4. Conclusion

Viscoelastic phase separation, a phenomenon unique to polymer and colloidal systems displaying dynamic asymmetry between components, is a powerful thermodynamic process to form interconnected network structures. As revealed by a combination of rheological characterization and temperature-dependent advanced microscopy methods, we have discovered a pathway to arrest viscoelastic phase separation in RR-P3HT solutions at early stages via interchain crystallization in order to form semicrystalline gel networks with hierarchical porosity. RR-P3HT network strength was insensitive to the presence of both PCBM and regiorandom P3HT, as monitored by oscillatory shear measurements. The abiilty to form the porous network in the presence of a second component will have a broad impact in a variety of fields, including optoelectronics. By affording continuous pathways for charge transport and large interfacial area with the electron-accepting PCBM phase, the presence of these RR-P3HT networks in the photoactive layers of OSCs was demonstrated to enhance solar conversion efficiency from $2.7 \pm 0.2\%$ to $3.9 \pm 0.2\%$. Significantly, these networks were formed in solution *prior to photoactive layer deposition* in a highly reproducible manner. Crystallization-arrested viscoelastic phase separation thus presents a promising strategy to achieve optimal photoactive layer morphologies by controlling the structure of polymer-small molecule solutions prior to deposition. Furthermore, this processing method is compatible with large-scale manufacturing methods to continuously process OSC photoactive layers from solution, which will be critical in advancing this technology towards commercialization.

## Supporting Information

The Supporting Information is available free of charge on the ACS Publications website:

Methods, design and supporting figures of the characterization data (PDF)

## Corresponding Author

*Stephanie S. Lee, E-mail: stephanie.lee@stevens.edu

## ACKNOWLEDGMENT

This material is based upon work supported by the National Science Foundation under Grant No. 1635284. The authors are also grateful for noteworthy support from PSEG to advance energy innovation at Stevens. Portions of the research used microscopy resources within the Laboratory for Multiscale Imaging at Stevens Institute of Technology.

**Crystallization-Arrested Viscoelastic Phase Separation in Semiconducting Polymer Gels**

**Supporting Information**

*Jing He, Xiaoqing Kong, Yuhao Wang, Michael Delaney, Dilhan M. Kalyon and Stephanie S. Lee**

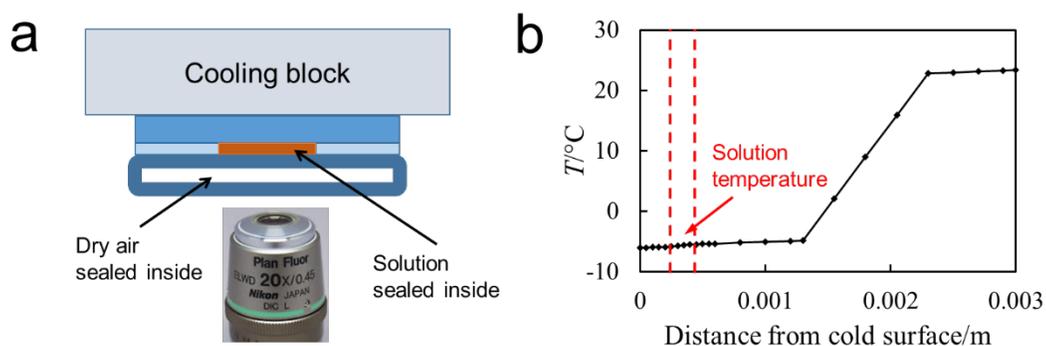

**Figure S1**. a) Setup of temperature-variable confocal laser scanning microscopy experiments. b) Temperature distribution obtained from temperature and heat flux sensor measurement.

**Small amplitude oscillatory shear**

During oscillatory shearing the shear strain, $\gamma$, varies sinusoidally with time, $t$, at a frequency of $\varpi$, i.e., $\gamma(t) = \gamma^0 \sin(\varpi t)$ where $\gamma^0$ is the strain amplitude. The shear stress, $\tau(t)$ response of the fluid to the imposed oscillatory deformation consists of two contributions associated with the energy stored as elastic energy and energy dissipated as heat:

$$\tau(t) = G'(\varpi)\gamma^0 \sin(\varpi t) + G''(\varpi)\gamma^0 \cos(\varpi t)$$

The storage modulus, $G'(\varpi)$ and the loss modulus, $G''(\varpi)$ also define the magnitude of complex viscosity, $|\eta^*| = \sqrt{(G'/\varpi)^2 + (G''/\varpi)^2}$, and $\tan\delta = G''/G'$. In the linear viscoelastic region all dynamic properties are independent of the strain amplitude, $\gamma^0$.

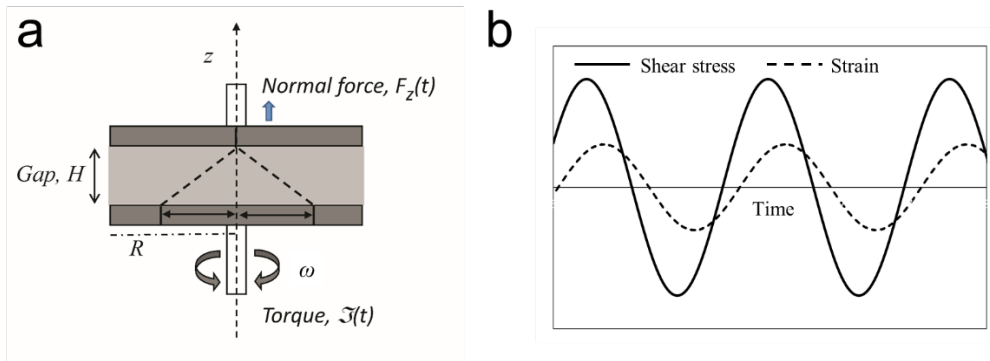

**Figure S2**. a) Schematic of parallel disk rheometer during small amplitude oscillatory shear characterization. b) Graph of the sinusoidal strain applied to samples and its corresponding stress versus time.

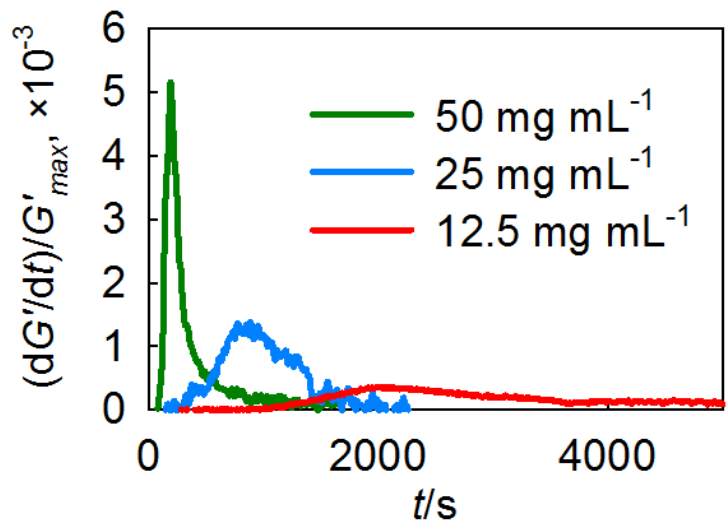

**Figure S3**. First time derivative of the storage modulus, $dG'/dt$, normalized to the value of steady state storage modulus $G'_{max}$, versus time for different concentrations of RR-P3HT solutions at -5 °C. The time where maximum value of $dG'/dt$ was observed was denoted as $t_c$ to track the gelation rate.

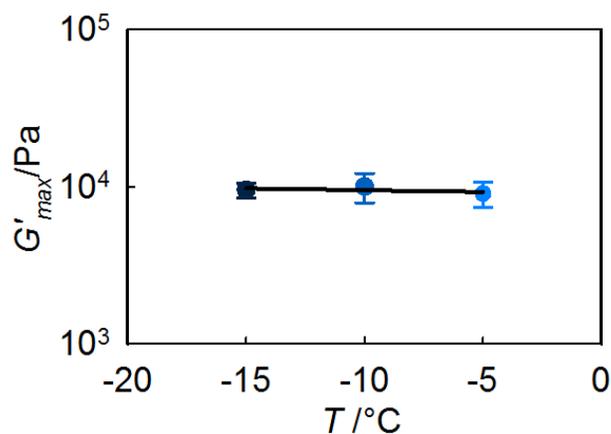

**Figure S4**. The steady-state storage modulus, $G'_{max}$, of a 25 mg/mL RR-P3HT solution versus temperature.

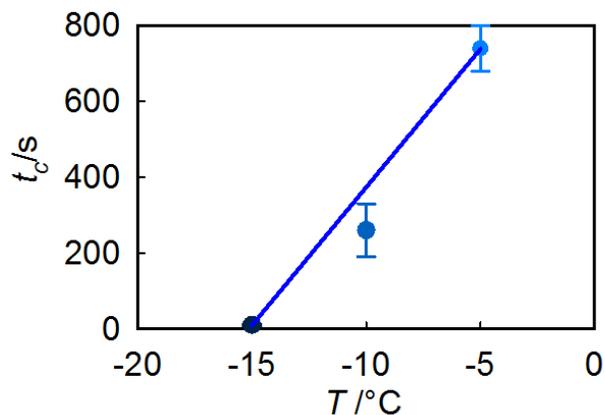

**Figure S5**. The characteristic time of gelation, $t_c$, of a 25 mg/mL RR-P3HT solution versus temperature.

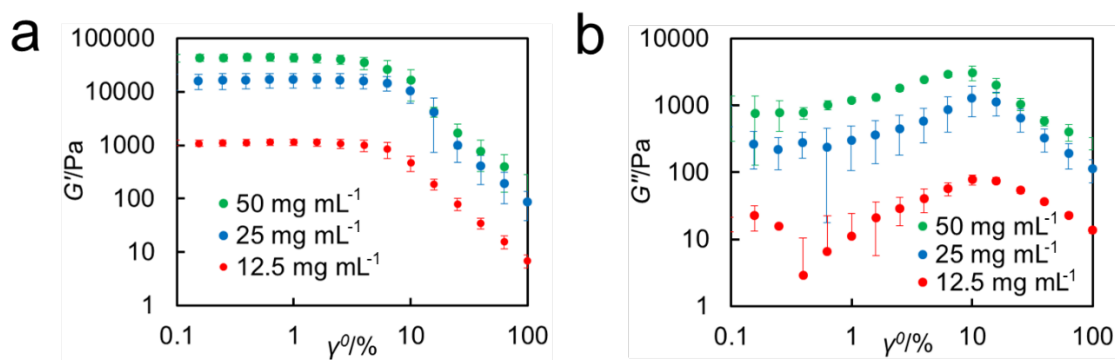

**Figure S6**. a) Storage modulus, $G'$ and b) loss modulus, $G''$, versus strain amplitude, $\gamma^0$, at a frequency of 1 rad/s of RR-P3HT solutions in o-DCB with concentrations of 12.5, 25, and 50 mg/mL after reaching steady-state at -5 °C.

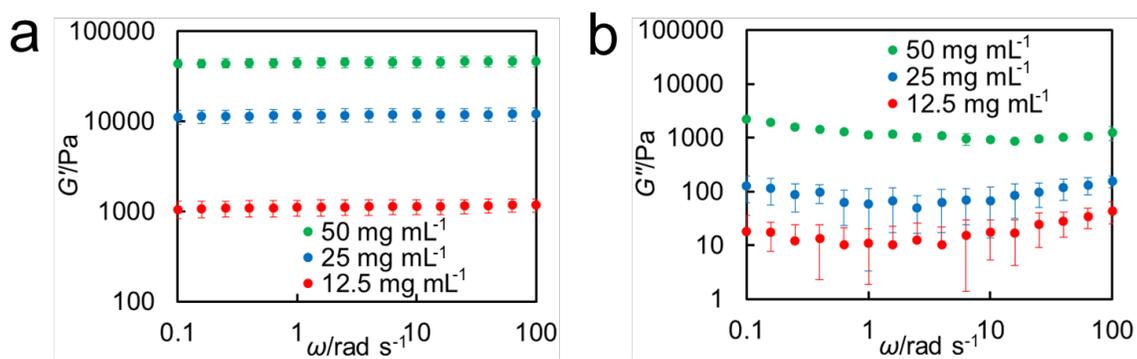

**Figure S7**. a) Storage modulus, $G'$, b) Loss modulus, $G''$, versus frequency, $\varpi$ at a strain amplitude, $\gamma^0$, of 1% of RR-P3HT solutions in o-DCB with concentrations of 12.5, 25, and 50 mg/mL after reaching steady-state at -5 °C.

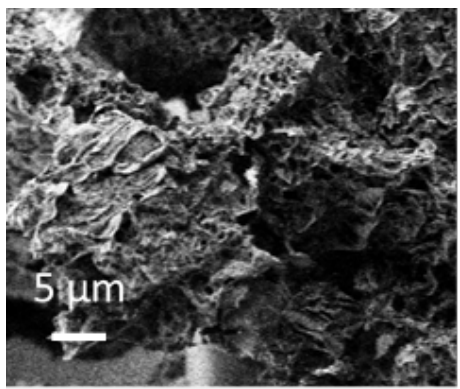

**Figure S8**. Cryo-SEM image of a RR-P3HT gel in o-DCB after overnight sublimation to remove all of the solvent.